# Thermodynamic Properties of Ferromagnetic Mott- Insulators GaV$_4$S$_8$

C. S. Yadav [a], A. K. Nigam[b] and A. K. Rastogi[a]

[a] *School of Physical sciences, Jawaharlal Nehru University, Newdelhi-110067,India*
[b] *Tata Institute of Fundamental Research, Homi Bhabha Road, Colaba, Mumbai-400005,India*

**Abstract**

We present the results of the magnetic and specific heat measurements on V$_4$ tetrahedral-cluster compound GaV$_4$S$_8$ between 2 to 300K. We find two transitions related to a structural change at 42K followed by ferromagnetic order at 12K on cooling. Remarkably similar properties were previously reported for the cluster compounds of Mo$_4$. These compounds show an extremely high density of low energy excitations in their electronic properties. We explain this behavior in a cluster compound as due to the reduction of coulomb repulsion among electrons that occupy highly degenerate orbits of different clusters.
© 2001 Elsevier Science. All rights reserved

In common with the compounds of Nb, Ta and Mo , the V- cluster compound GaV$_4$S$_8$ is an insulators due to large inter-cluster distances. Unlike Mott- insulators (NiO etc.) these compounds show the absence of a correlation gap in their electronic excitation spectrum. The electronic transport is here by hopping of localized carriers at the Fermi energy. The V and Mo – cluster compounds show an exceptional ferromagnetic order and their thermodynamic properties indicate a curious mixture of the electrons occupying a band of energies and the localized spin fluctuations on the individual clusters [1,2]. The specific heat is also extremely large due to high density of the spin/charge excitations at low temperatures [3].

In figure 1, we show the temperature dependence of the magnetic properties of pure GaV$_4$S$_8$ and also after substitution of Ga by 5-10% Cu. At high temperatures the susceptibility, shown in the inset as $\chi^{-1}(T)$, can be fitted to a sum of a curie term ( with $\mu_{eff}$ of 1.38 $\mu_B$ on each V$_4$-cluster), and a weakly temperature dependent van Vleck term $\chi_{v.v}$ equal to $7\times10^{-4}$ (emu/mole) . These values clearly indicate that the magnetism here is due to unpaired electrons which occupy closely spaced orbitals a V$_4$ – cluster.

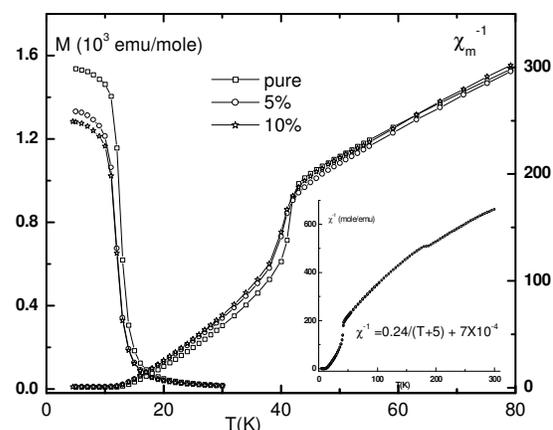

Fig. 1: Magnetic properties of (Ga$_x$Cu$_{1-x}$)V$_4$S$_8$ showing structural transitions at 42K and ferromagnetism at 12.1K

There is a first order cubic to rhombohedral structural transition on cooling to 42K , where subtle changes in the electronic structure of the cluster units have been reported[4,5]. The structural transition causes a discontinuous change in the susceptibility and a dramatic increase in the magnetic interaction in the low temperature phase. $GaV_4S_8$ becomes ferromagnetic below12K. The magnetization measurement at 4.2K and 6Tesla field gave magnetic moment of 0.72 $\mu_B$ / cluster. Similar succession of transitions has been seen in almost all the compounds of $V_4$ and $Mo_4$ [3].

The most remarkable magnetic property is however, for the dependence of the magnetic susceptibility on temperature.. As shown in figure 2, $\chi^{-1}$ closely follows a quadratic dependence on temperature -- $(2\chi_o S)^{-1} [(T/T_C)^2 - 1]$, extending over a large temperature range in the paramagnetic phase. This dependence is same as expected (but never before observed) in the Stoner- Wohlfarth model of itinerant electron magnetism [3]. Moreover, we find that the extrapolated value of $T_C$ (=12.4K ) obtained from the $\chi^{-1}(T^2)$ plot is remarkably close to the critical value of 12.1K seen in our specific heat results.

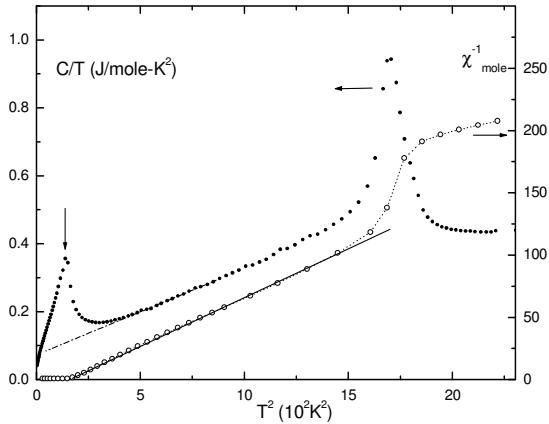

Fig. 2: Susceptibility ( $\chi$ ) and specific heat( C ) results of $GaV_4S_8$ plotted as $\chi^{-1}$ and C/T versus $T^2$

In figure 2, we have also plotted the specific heat results of $GaV_4S_8$ showing the two transitions due to ferromagnetic order and structural change. From the straight line plots of C/T versus $T^2$, we deduce an electronic term $\gamma_{el}T$, with $\gamma_{el}$ = 75mJ/mole-$K^2$ in its paramagnetic phase and a phonon contribution giving a value of 446 K for the Debye temperature $\theta_D$. These values can be compared with the values of 190 mJ/mole-$K^2$ and 390K respectively for the $\gamma_{el}$ and $\theta_D$ obtained for $GaMo_4S_8$ [3].

The results shown in figure 2 are extraordinary and clearly mark these cluster compounds as a distinct family of Mott insulators. We believe that an electronic contribution in the specific heat and the quadratic temperature dependence found in the paramagnetic susceptibility are closely related to the electrons occupying band of orbitals localized on the clusters. The electronic correlations are considerably reduced and on site Hunds rule coupling gives ferromagnetism .The role of structural transition for the ferromagnetism is the subject of detailed investigation.

CSY acknowledges CSIR, India for the financial grant.